\documentclass[a4paper,11pt]{article}
\pdfoutput=1

\usepackage{jheppub} 

\usepackage{lmodern}

\usepackage{booktabs}
\usepackage{subcaption}
\newsavebox{\largestimage}
         
\usepackage{float}         
\allowdisplaybreaks

\usepackage[T1]{fontenc} 

\usepackage{tensor}
\usepackage{tikz}

\newcommand{\mr}[1]{\mathrm{#1}}

\newcommand{\ms}[1]{\mathsf{#1}}
\newcommand{\mc}[1]{\mathcal{#1}}
\newcommand{\mb}[1]{\mathbb{#1}}

\newcommand{\f}[2]{\frac{#1}{#2}}

\newcommand{\rEs}[6]{%
\tensor[_{#1}]{E}{_{#2}}%
\left(\genfrac..{0pt}{}{#3}{#4}\middle|#5\middle|#6%
\right)%
}

\let\l\relax
\let\r\relax
\let\d\relax
\newcommand{\l}{\left}
\newcommand{\r}{\right}
\newcommand{\d}{\mr{d}}
\newcommand{\dts}{\!\ifmmode\mathinner{\ldotp\kern-0.2em\ldotp\kern-0.2em\ldotp}\else.\kern-0.13em.\kern-0.13em.\fi\!}

\title{Elliptic Genus Derivation of 4d Holomorphic Blocks}

\author{Matteo Poggi}
\emailAdd{mpoggi@sissa.it}
\preprint{SISSA~55/2017/FISI}

\affiliation{International School of Advanced Studies (SISSA/ISAS) and INFN Sezione di Trieste\\
via Bonomea 265, I-34136 Trieste, Italy}

\abstract{We study elliptic vortices on $\mb{C}\times T^2$ by considering the 2d quiver gauge theory describing their moduli spaces. The elliptic genus of these moduli spaces is the elliptic version of vortex partition function of the 4d theory. We focus on two examples: the first is a $\mc{N}=1$, $\mr{U}(N)$ gauge theory with fundamental and anti-fundamental matter; the second is a $\mc{N}=2$, $\mr{U}(N)$ gauge theory with matter in the fundamental representation. The results are instances of 4d ``holomorphic blocks'' into which partition functions on more complicated surfaces factorize. They can also be interpreted as free-field representations of elliptic Virasoro algebrae.}

\begin{document}
 \maketitle
 
\section{Elliptic vortices: Introduction}

The non perturbative analysis of quantum field theories is one of the most interesting problems in contemporary 
theoretical high energy physics.
In the case of supersymmetric theories, some techniques have been worked out in the last two decades to calculate
their exact features.
When the quantum field theory is formulated on a non trivial manifold, the gluing of the space time manifold in local patches 
should reflect in a factorization property of the partition function in local blocks.
In particular, the supersymmetric quantum field theory can be studied on a Riemannian manifold with isometries to obtain its supersymmetric partition function by exploiting the higher degree of symmetry of the problem.
This has been done for geometries obtained by products of spheres, tori and lines in different combinations, starting from \cite{pestun} (see \cite{locrev} for a comprehensive review).

An intriguing aspect of these constructions, is that the structure of the Riemannian manifold in patches reflects in factorization properties
of the partition function in partial blocks. A simple example of this phenomenon is the decomposition of two-sphere SUSY partition functions 
in North and South patches contributions, as in \cite{bc, gd}, in terms of vortex/anti-vortex contributions \cite{vortex, zhao, shad, dgh}.
Because of supersymmetry, these 2d blocks are holomorphic/anti-holomorphic functions of a complex combination of the vortex counting parameter and the Fayet-Iliopoulos one, making the North and South patches decomposition of the SUSY theory an holomorphic factorization.
This is clearly a property implied by the BPS saturation of vortex/anti-vortex configurations.

A particularly interesting phenomenon of this kind in higher dimensions has been observed in \cite{hf3, hf4, bdp}, where the factorization of supersymmetric partition functions on squashed three-spheres and on their products with circles has been noticed to happen in terms of 3d and 4d ``holomorphic blocks''. Such factorization has also been derived using Higgs branch localization in \cite{bp,fmy}. Other examples of factorization in 4d can be found in \cite{pe,yo}.
While the former were identified with vortex particles blocks on ${\mathbb C}\times S^1$, the latter await a first principle computation as 
elliptic vortices on ${\mathbb C}\times T^2$, while their description in terms of elliptic hypergeometric functions \cite{spiridonov} is given by holomorphic factorization.

The results of this paper, when specialized to the appropriate case, provide such a first principle evaluation. This is done by computing the SUSY gauge theory partition function on ${\mathbb C}\times T^2$ by resumming its expansion in rotational modes on the complex plane (elliptic vortices), each term being the elliptic genus of the corresponding vortex moduli space. This is computed by adapting the techniques of \cite{bh1, bh} to a proper supersymmetric gauge theory encoding 
such a vortex moduli space in its low energy $\sigma$-model phase. 
The calculations of the proper elliptic genera is performed by extracting the relevant Jeffrey-Kirwan residues 
of the BPS one-loop determinants of the 2d auxiliary theory.
Resumming all the partial vortex contributions, this is shown to exactly reproduce the 4d holomorphic blocks of \cite{hf4}.

Another motivation for this work is to understand the algebraic structure of BPS vacua of such theories. Indeed, it is by now well known that Virasoro algebra and its generalization to W-algebrae
acts on the moduli space of instantons in four-dimensions \cite{AGT,BT, wyll, mm}. Vortices are the two-dimensional analogue of instantons and indeed their moduli spaces can be obtained as special Lagrangian submanifolds
of the instanton moduli space \cite{hanany-tong,vortex}. It is thus interesting to investigate and unveil the algebrae acting on their equivariant cohomology spaces. On the mathematical side, a natural related question arises concerning the algebrae which can be realized
on the moduli spaces of sheaves over the cotangent bundle of a Riemann surface $T^*\Sigma$. This paper indicates a candidate algebra for the simplest case of the two-torus $\Sigma=T^2$.
Indeed we will show that the resummed elliptic vortex partition function coincides with the appropriate free-field correlator of elliptic Virasoro algebra as studied in \cite{nieri}.

A specialization of the vortex partition functions we analyze in this paper can be also obtained from a six-dimensional gauge theory on $\mathbb{R}^4\times T^2$
in presence of a codimension two defect along an $\mathbb{R}^2\times T^2$ \cite{nieri,zabzine}. Indeed, this gives rise to a coupled 6d-4d system which reduces to the elliptic genus computation in the decoupling limit of the 6d dynamics.

\section{Elliptic genus of vortex moduli spaces}

In this Section we perform a thorough study of the elliptic genus of moduli spaces of vortices.
We consider two basic cases. The first is a $\mc{N}=(0,2)$ quiver gauge theory as depicted in fig.~\ref{fig:1}. This describes
the moduli space of $k$ vortices in a four-dimensional $\mathcal{N}=1$ supersymmetric gauge theory with $U(N)$
gauge group, $N_\mathrm{F}$ fundamental chirals and $\tilde{N_\mathrm{F}}$ anti-fundamental chirals, \cite{hanany-tong}. As we will show in the following,
anomaly cancellation in the 2d theory requires $\tilde{N_\mathrm{F}}=N_\mathrm{F}$, thus reflecting the anomaly cancellation condition of the parent
four-dimensional theory. The second is a $\mc{N}=(2,2)$ quiver gauge theory as depicted in fig.~\ref{fig:2}, describing the moduli space of $k$ vortices
in $\mathcal{N}=2$, $U(N)$ gauge theory with $N_\mathrm{F}$ hypermultiplets in the fundamental \cite{hanany-tong}.
\subsection{$\mc{N}=(0,2)$ theories}
\begin{figure}[h]
  \centering
    \includegraphics{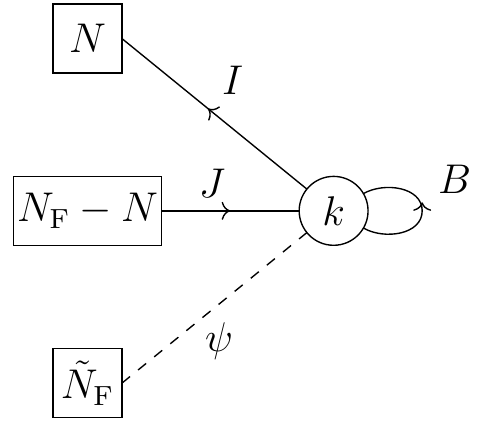}
  \caption{$\mc{N}=(0,2)$ quiver gauge theory: it has\protect\footnotemark $\mr{U}(k)$ vector multiplet, $I$, $J$ and $B$ chiral multiplets and $\psi$ Fermi multiplet. It describes the moduli space of vortices of 4d, $\mc{N}=1$, $\mr{U}(N)$ gauge theory with $N_{\mr{F}}$ fundamental chirals and $\tilde{N}_{\mr{F}}$ antifundamental chirals. }
  \label{fig:1}
\end{figure}

Our aim is to compute the elliptic genus for the quiver in fig.\ref{fig:1}, following \cite{bh1,bh}.
The holonomies that are switched on are listed in tab.~\ref{tab:20}.
\footnotetext{All the multiplets here are $\mc{N}=(0,2)$ multiplets.}

\begin{table}[h]
 \centering
 \begin{tabular}{lc}
  \toprule
  Group & Fugacity\\
  \midrule
  $\mr{U}(k)$ & $u_i$\\
  $\mr{U}(N)$ & $z_\alpha$\\
  $\mr{U}(N_\mr{F}-N)$ & $\mu_A$\\
  $\mr{U}(\tilde{N}_\mr{F})$ & $\nu_I$\\
  $\mr{U}(1)_B$ & $-\epsilon$\\
  \bottomrule
 \end{tabular}
 \caption{Holonomies and Fugacities of the $\mc{N}=(0,2)$ case.}\label{tab:20}
\end{table}
\noindent Let us notice that the flavor group of the theory is actually $\mr{S}(\mr{U}(N) \times \mr{U}(N_{\mr{F}}-N) \times \mr{U}(\tilde{N}_{\mr{F}}))$, since the overall phase of $\mr{U}(N) \times \mr{U}(N_{\mr{F}}-N) \times \mr{U}(\tilde{N}_{\mr{F}})$ belongs to the gauge group $\mr{U}(k)$. This means that the fugacities in tab.~\ref{tab:20} are not independent but subject to the constraint:
 \begin{equation}\label{eq:anomaly_canc}
 \sum_{\alpha=1}^Nz_\alpha + \sum_{A=1}^{N_{\mr{F}}-N}\mu_A - \sum_{I=1}^{\tilde{N}_{\mr{F}}}\nu_I=0.
\end{equation}
The elliptic genus therefore reads:
\begin{equation}\label{eq:partfunc20}
 \mc{Z}_{N,k}^{(0,2)}(\vec{z},\vec{\mu},\vec{\nu},\epsilon,\tau) = \int Z_\mr{vec}(\vec{u},\tau) Z_B(\vec{u},\epsilon,\tau) Z_I(\vec{u},\vec{z},\tau) Z_J(\vec{u},\vec{\mu},\tau) Z_\psi(\vec{u},\vec{\nu},\tau),
\end{equation}
where the contour of integration is specified by Jeffrey-Kirwan prescription (see sec.~\ref{sss:polology1} for further details), and the contributions of the multiplets are respectively:
\begin{align}
    Z_\mr{vec}(\vec{u},\tau) &= \label{eq:Z_vec}
    \f{1}{k!}\l(\f{2\pi\eta^2(q)}{i}\r)^k 
    \prod_{\substack{r,s=1\\r\neq s}}^k i\f{\theta_1(\tau | u_r - u_s)}{\eta(q)}
    \prod_{r=1}^k \d u_r, \\
    Z_B(\vec{u},\epsilon,\tau) &=
    \prod_{r,s=1}^k i\f{\eta(q)}{\theta_1(\tau | u_r - u_s - \epsilon)},\\
    Z_I(\vec{u},\vec{z},\tau) &=
    \prod_{r=1}^k \prod_{\alpha=1}^N i\f{\eta(q)}{\theta_1(\tau | u_r - z_\alpha)},\\
    Z_J(\vec{u},\vec{\mu},\tau) &=
    \prod_{r=1}^k \prod_{A=1}^{N_\mr{F}-N} i\f{\eta(q)}{\theta_1(\tau | -u_r + \mu_A)},\\
    Z_\psi(\vec{u},\vec{\nu},\tau) &=
    \prod_{i=r}^k \prod_{I=1}^{\tilde{N}_\mr{F}} i\f{\theta_1(\tau | u_r - \nu_I)}{\eta(q)}.
\end{align}
\subsubsection{Anomaly Cancellation}
Since our 2d theory is manifestly chiral, we have to find out under which conditions gauge anomalies cancel. A very instructive way to find such conditions is to impose the double periodicity of the integrand of the partition function eq. \eqref{eq:partfunc20}. Such property is \emph{not} trivially enjoyed since eq. \eqref{eq:thetatrasf} shows us that $\theta_1$ is just \emph{quasiperiodic} under a shift which is an integer times the modulus of the torus. Let us study the behaviour of the integrand in eq. \eqref{eq:partfunc20} under the shift: $u_i \mapsto u_i + a + b\tau$ ($a,b\in\mb{Z}$):
\begin{itemize}
 \item $Z_{\mr{vec}}$ and $Z_B$ are left unchanged;
 \item $Z_I \mapsto 
        (-1)^{kN(a+b)}
        e^{i\pi k N  b^2 \tau}
        e^{2\pi iN b \sum_{j=1}^k u_j}
        e^{-2\pi i k b\sum_{\alpha=1}^Nz_\alpha}
        Z_I$;
 \item $Z_J \mapsto 
        (-1)^{k(N_{\mr{F}}-N)(a+b)}
        e^{i\pi  k(N_{\mr{F}}-N)b^2\tau}
        e^{2\pi i (N_{\mr{F}}-N)b\sum_{j=1}^ku_j}
        e^{-2\pi i k b \sum_{A=1}^{N_{\mr{F}}-N}\mu_A}
        Z_J$;
 \item $Z_\psi \mapsto
        (-1)^{-k\tilde{N}_{\mr{F}}(a+b)}
        e^{-i\pi i k\tilde{N}_{\mr{F}} b^2 \tau}
        e^{-2\pi i \tilde{N}_{\mr{F}} b \sum_{j=1}^{k} u_i}
        e^{2\pi i k b \sum_{I=1}^{\tilde{N}_{\mr{F}}}\nu_I}
        Z_\psi$.
\end{itemize}
Combining all these contributions together and imposing shift invariance we get:
\begin{multline}
 (-1)^{k(N_{\mr{F}}-\tilde{N}_{\mr{F}})(a+b)}
 e^{i\pi k (N_{\mr{F}}-\tilde{N}_{\mr{F}})b^2 \tau}
 e^{2\pi i (N_{\mr{F}}-\tilde{N}_{\mr{F}})b \sum_{j=1}^k u_j}\times\\
 \times
 e^{-2\pi i k b \l(\sum_{\alpha=1}^Nz_\alpha + \sum_{A=1}^{N_{\mr{F}}-N}\mu_A - \sum_{I=1}^{\tilde{N}_{\mr{F}}}\nu_I \r)}
 =1.
\end{multline}
The last exponential is one thanks to eq. \eqref{eq:anomaly_canc}, thus the anomaly cancels iff:
\begin{equation}
 N_{\mr{F}} = \tilde{N}_{\mr{F}}.
\end{equation}
\subsubsection{Polology} \label{sss:polology1}
In order to evaluate the integral \eqref{eq:partfunc20} we will use the prescription given by Jeffrey and Kirwan (see \cite{jk} for the mathematical formulation, \cite{w}, for the conjecture from which it originates, \cite{bh1, bh}, for its application to the elliptic genus). It is a procedure to select which poles one should take the residue at.
There are three sources of poles, that is $Z_I$, $Z_J$ and $Z_B$ in \eqref{eq:partfunc20}. The singular hyperplanes they lie on are:
\begin{equation}\label{eq:singular_hyperplanes}
 H_{B;r,s} = \l\{u_r - u_s - \epsilon = 0 \r\},
 \qquad\qquad
 H_{I;r,\alpha} = \l\{ u_r = z_\alpha \r\},
 \qquad\qquad
 H_{J;r,A} = \l\{ u_r = \mu_A \r\},
\end{equation}
whose charge are:
\begin{equation}
 \vec{h}_{B;r,s}= (0, \dts, \underbrace{1}_r, \dts, \underbrace{-1}_s, \dts, 0),
 \qquad
 \vec{h}_{I;r,\alpha} = (0, \dts, \underbrace{1}_r, \dts, 0),
 \qquad
 \vec{h}_{J;r,A} = (0, \dts, \underbrace{-1}_r, \dts, 0).
\end{equation}

Let us now describe the procedure to find all poles that give a contribution to the elliptic genus at order $k$ (where $k$ is the number of vortices in the 4d theory). First, we have to find all the intersection points of $k$ hyperplanes described by eqs. \eqref{eq:singular_hyperplanes}; then we have to select the poles corresponding to the charge matrices whose cone contains a given vector $\vec{\eta}$. In the following\footnote{The final result does \emph{not} depend on the choice of that $\vec{\eta}$, however, as pointed out in \cite{bh}, if one does not take it parallel to the Fayet-Iliopoulos term, an extra contribution may arise from infinity.} we will take $\vec{\eta}=(1,\dts ,1)$.

In the first step we have to select all the possible groups of $k$ equations among \eqref{eq:singular_hyperplanes} that have a \emph{unique} solution. This corresponds to look for a solution of the following system of linear equations:
\begin{equation}
 \begin{pmatrix}
  \vec{h}_1\\
  \vdots\\
  \vec{h}_k
 \end{pmatrix}
 \begin{pmatrix}
  u_1\\
  \vdots\\
  u_k
 \end{pmatrix}
 =
 \begin{pmatrix}
  c_1\\
  \vdots\\
  c_k
 \end{pmatrix}
\end{equation}
where $\vec{h}_r$ are chosen among eqs. \eqref{eq:singular_hyperplanes}, and $c_r$ is $\epsilon$ if the corresponding $\vec{h}_i$ is of the type $H_B$; $z_\alpha$, for some $\alpha$, if the corresponding corresponding $\vec{h}_r$ is of the type $H_I$; and $-\mu_A$, for some $A$, if the corresponding $\vec{h}_r$ is of the type $H_J$. We notice that if we take all $\vec{h}$'s of the type $H_B$, the coefficient matrix of the system is singular, since the sum of every row is zero, and hence such system doesn't have a solution. Thus, we must select \emph{at least} one equation coming from $H_I$ or $H_J$. These equations will set some of the $u$'s equal to fugacities $z_\alpha$ or $\mu_A$: let us call such $u$'s in the following way:
\begin{equation}\label{eq:first_us}
 u_{\alpha,1} = z_\alpha,
 \qquad\qquad\qquad\qquad
 u_{A,1} = \mu_A.
\end{equation}
Then, using equation of the type $H_B$, one can find all others $u$'s starting from eqs. \eqref{eq:first_us}, so that every solution is of the form:
\begin{equation}
 u_{\alpha,r_\alpha} = z_\alpha + (r_\alpha - 1)\epsilon,
 \qquad\qquad\qquad
 u_{A,r_A} = \mu_A + (r_A - 1)\epsilon,
\end{equation}
with $r_\alpha \in[r_\alpha^{(\mr{min})},r_\alpha^{(\mr{max})}]$, $r_A\in [r_A^{(\mr{min})},r_A^{(\mr{max})}]$, with the condition $k=\sum_{\alpha}k_\alpha + \sum_A k_A$, where $k_\alpha = r_\alpha^{(\mr{max})}-r_\alpha^{(\mr{min})}$, $k_A= r_A^{(\mr{max})}-r_A^{(\mr{min})}$. We remark here that all the $u$s must be different, otherwise the residue would vanish because of the $\theta_1(\tau | u_r - u_s)$ in the numerator.

Next, we have to check whether $\vec{\eta}$ lies in the cone generated by $\vec{h}_1,\dts ,\vec{h}_k$. This is equivalent to check if
\begin{equation}\label{eq:JKs}
 \begin{pmatrix}
  \vec{h}_1^\ms{T}
  &
  \dots
  &
  \vec{h}_k^\ms{T}
 \end{pmatrix}
 \begin{pmatrix}
  \beta_1\\
  \vdots\\
  \beta_k
 \end{pmatrix}
 =
 \vec{\eta}^\ms{T}
\end{equation}
has a solution for $\beta_i>0$. It is easy to see that the prototype of the coefficient matrix in eq. \eqref{eq:JKs} that gives a solution with positive $\beta$'s is a block-diagonal matrix (one block for each $\alpha$ or $A$) whose blocks are of the form:
\begin{equation}\label{eq:JKmatrix}
 \begin{pmatrix}
  1 & -1 & * & * & \dots & *\\
  0 &  1 & * & * & \dots & *\\
  0 &  0 & 1 & * & \dots & *\\
  \vdots & \vdots & \vdots & \vdots & \ddots & \vdots\\
  0 &  0 & 0 & 0 & \dots & 1
 \end{pmatrix},
\end{equation}
where each $*$ can be either $0$ or $-1$ in such a way every column is a charge vector $\vec{h}$. From this form we can read off which hyperplanes give a JK pole: they are $H_{I;r,\alpha}$ (first column) and $H_{B;r,s}$ with $r>s$ (remaining columns). Other hyperplanes give JK-excluded poles since their charge vectors are related to those appearing in \eqref{eq:JKmatrix} by a sign flip. This leads to a flip in the corresponding $\beta$ which becomes negative. 

Thus, we conclude that the poles at which we have to evaluate the residue are just:
\begin{equation}\label{eq:poles}
 u_{\alpha,r_\alpha} = z_\alpha + (r_\alpha - 1)\epsilon,
\end{equation}
with $r_\alpha \in [1, k_\alpha]$, where $k=\sum_\alpha k_\alpha$. These poles can be represented as a collection of ``colored'' stripes of boxes: each color represents a different $\alpha$. Of course, given a set of poles $k$ that contribute to the elliptic genus, any of its $k!$ permutation will contribute as well. We see that this multiplicity factor cancels the order of the Weyl group in eq. \eqref{eq:Z_vec}.

\subsubsection{Computation}
Now we have to compute residues at the poles found in the preceding paragraph: this is pretty simple since we are dealing with simple poles. We have just to use eq. \eqref{eq:theta_res} with $a=b=1$ for the $\theta_1$ giving the pole, and to evaluate the other non singular factors at that point. Since we labelled poles with two indices (eq. \eqref{eq:poles}) it will be convenient to rearrange products as $\prod_{i=1}^k = \prod_{\alpha=1}^N\prod_{r_\alpha=1}^{k_\alpha}$. In this way it's easy to write the contribution of every multiplet:
\begin{align}
 \mc{Z}_{\mr{vec}} &= 
        \l(\f{2\pi\eta^2(q)}{i}\r)^k
        \prod_{\substack{\alpha,\beta=1\\ \alpha\neq\beta}}^N\prod_{r_\alpha=1}^{k_\alpha}\prod_{s_\beta=1}^{k_\beta}
        i\f{\theta_1(\tau | z_{\alpha\beta}+(r_\alpha - s_\beta)\epsilon)}{\eta(q)}\times\nonumber\\
        &\hspace{2cm}\times
        \prod_{\alpha=1}^N \prod_{\substack{r_\alpha, s_\alpha=1 \\ r_\alpha\neq s_\alpha}}^{k_\alpha}i\f{\theta_1(\tau | (r_\alpha - s_\alpha)\epsilon)}{\eta(q)},
        \\
 \mc{Z}_B &= 
        \prod_{\substack{\alpha,\beta=1\\ \alpha\neq\beta}}^N\prod_{r_\alpha=1}^{k_\alpha}\prod_{s_\beta=1}^{k_\beta}
        i\frac{\eta(q)}{\theta_1(\tau | z_{\alpha\beta}+(r_\alpha-s_\beta -1)\epsilon)}
        \times\nonumber\\
        &\hspace{2cm}\times
        \prod_{\alpha=1}^N \l(\f{1}{2\pi\eta^3(q)}\r)^{k_\alpha -1}\prod_{\substack{r_\alpha, s_\alpha=1 \\ r_\alpha\neq s_\alpha+1}}^{k_\alpha}i\f{\eta(q)}{\theta_1(\tau | (r_\alpha-s_\alpha-1)\epsilon)},\\
  \mc{Z}_I & = 
        \prod_{\substack{\alpha,\beta=1\\ \alpha\neq\beta}}^N \prod_{r_\alpha=1}^{k_\alpha}
        i\frac{\eta(q)}{\theta_1(\tau | z_{\alpha\beta} + (r_\alpha -1)\epsilon)}
        \times 
        \prod_{\alpha=1}^N \l(\f{1}{2\pi\eta^3(q)} \r)
        \prod_{r_\alpha=2}^{k_\alpha}i\f{\eta(q)}{\theta_1(\tau | (r_\alpha -1)\epsilon)},
        \\
  \mc{Z}_J & = 
        \prod_{\alpha=1}^N \prod_{A=1}^{N_{\mr{F}}-N}\prod_{r_\alpha=1}^{k_\alpha}i\f{\eta(q)}{\theta_1(\tau | \mu_A - z_\alpha - (r_\alpha -1)\epsilon)},\\
  \mc{Z}_\psi & = 
        \prod_{\alpha=1}^N \prod_{I=1}^{N_{\mr{F}}} \prod_{r_\alpha=1}^{k_\alpha}
        i\f{\theta_1(\tau | z_\alpha - \nu_I + (r_\alpha-1)\epsilon)}{\eta(q)},
\end{align}
where $z_{\alpha\beta}= z_\alpha - z_\beta$. Now it is simply a matter of collecting and rearranging factors. According to:
\begin{equation} \label{eq:simp1}
 \f{
 \prod_{\substack{i_\alpha, s_\alpha=1 \\ r_\alpha\neq s_\alpha}}^{k_\alpha}\theta_1(\tau | (r_\alpha-s_\alpha)\epsilon)}{\prod_{\substack{r_\alpha, s_\alpha=1 \\ r_\alpha\neq s_\alpha+1}}^{k_\alpha}\theta_1(\tau | (r_\alpha - s_\alpha -1)\epsilon)\times\prod_{r_\alpha=2}^{k_\alpha}\theta_1(\tau | (r_\alpha-1)\epsilon)}
 =
 \f{1}{\prod_{r_\alpha=1}^{k_\alpha}\theta_1(\tau |- r_\alpha\epsilon)},
\end{equation}
and:
\begin{multline} \label{eq:simp2}
 \f{\prod_{r_\alpha=1}^{k_\alpha}\prod_{s_\beta=1}^{k_\beta}\theta_1(\tau | z_{\alpha\beta}+(r_\alpha-s_\beta)\epsilon)}{\prod_{r_\alpha=1}^{k_\alpha}\prod_{s_\beta=1}^{k_\beta}\theta_1(\tau |z_{\alpha\beta}+(r_\alpha-s_\beta-1)\epsilon)\times \prod_{r_\alpha=1}^{k_\alpha}\theta_1(\tau | z_{\alpha\beta}+(s_\alpha-1)\epsilon)}
 =\\
 =\f{1}{\prod_{r_\alpha=1}^{k_\alpha}\theta_1(\tau | z_{\alpha\beta}+(r_\alpha-k_\beta -1)\epsilon)},
\end{multline}
we can write:
\begin{multline}\label{eq:fin20}
\mc{Z}_{\vec{k}}^{(0,2)}(\vec{z},\vec{\mu},\vec{\nu},\epsilon,\tau) =\\ (-1)^{k(N+1)}
        \prod_{\alpha=1}^N
        \f{
            \prod_{I=1}^{N_{\mr{F}}}
            \Theta_1(\tau, \epsilon | z_\alpha - \nu_I)_{k_\alpha}
        }{
            \Theta_1(\tau, \epsilon | \epsilon)_{k_\alpha}
            \prod_{\substack{\beta=1\\ \beta\neq\alpha}}^N
            \Theta_1(\tau, \epsilon | z_{\alpha\beta}- k_\beta\epsilon)_{k_\alpha}
            \times
            \prod_{A=1}^{N_\mr{F}-N}
            \Theta_1(\tau, \epsilon | z_\alpha - \mu_A)_{k_\alpha}
        },
\end{multline}
where the $\Theta_1$ function is a generalization of Pochhammer symbols and it is defined and studied in app. \ref{app:Theta}. This result is an elliptic generalization of $k$-vortex partition functions found in \cite{bc,gd, dgh, shad, vortex, fuji, yoshi}.
If we now define de grand-canonical partition function summing over all $N$-colored partitions of $k$, as:
\begin{equation}
 \mc{Z}_N^{(0,2)}(\vec{z}, \vec{\mu},\vec{\nu},\epsilon,\tau)=  \sum_{k_1=1}^\infty\dots\sum_{k_N=1}^\infty  \mc{Z}_{\vec{k}}^{(0,2)}(\vec{z}, \vec{\mu},\vec{\nu},\epsilon,\tau) \l[-\l(-e^{i\pi\epsilon} \r)^N z\r]^{|\vec{k}|},
\end{equation}
where $|\vec{k}|= \sum_{\alpha} k_i$. Applying eq. \eqref{eq:trick} to eq. \eqref{eq:fin20}, and then rewriting the prefactor as a differential operator, making use of eqs. \eqref{eq:baltheta} and \eqref{eq:anomaly_canc}, we can write:
\begin{equation}\label{eq:gc20}
 \mc{Z}_N^{(0,2)}(\vec{z}, \vec{\mu},\vec{\nu},\epsilon,\tau) = \mc{D}_N^{(0,2)} \prod_{\alpha=1}^N \rEs{N_{\mr{F}}}{N_{\mr{F}}-1}{\vec{A}_\alpha}{\vec{B}_\alpha, \vec{C}_\alpha}{\tau,\epsilon}{z_\alpha}
\end{equation}
in which we set:
\begin{align}
   \vec{A}_\alpha &= \l(z_\alpha - \nu_1,\dts, z_\alpha - \nu_{N_{\mr{F}}}\r);\\
   \vec{B}_\alpha &= \l(z_\alpha  - z_1,\dts, \widehat{z_\alpha - z_\alpha},\dts, z_\alpha - z_N\r);\\
   \vec{C}_\alpha &= \l(z_\alpha - \mu_A,\dts, z_\alpha - \mu_{N_{\mr{F}}-N} \r);\\
 \mc{D}_N^{(0,2)}
 &= \prod_{1\leq\alpha < \beta\leq N}e^{-i\pi \epsilon(z_\alpha\partial_{z_\alpha}-z_\beta \partial_{z_\beta})}\f{\theta(\tau | z_{\alpha\beta}+\epsilon(z_\alpha\partial_{z_\alpha}-z_\beta \partial_{z_\beta}))}{\theta(\tau | z_{\alpha\beta})},
\end{align}
where the wide hat means omission. We notice also that, using eq. \eqref{eq:derE} is possible to write the grand-partition function \eqref{eq:gc20} as a \emph{finite} combination of elliptic hypergeometric functions. Notice that in the case of $N=1$, the differential operator is not there and we recover the result of \cite{hf4}.

\subsection{$\mc{N}=(2,2)$ theories}
\begin{figure}[h]
  \centering
    \includegraphics{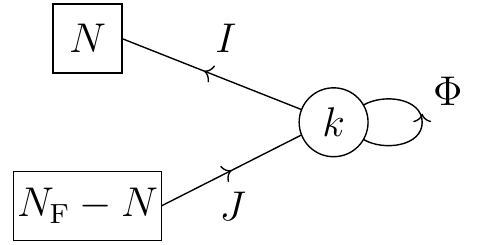}
  \caption{$\mc{N}=(2,2)$ quiver gauge theory: it has\protect\footnotemark $\mr{U}(k)$ vector multiplet, $I$, $J$ and $\Phi$ chiral multiplets. It describes the moduli space of vortices of 4d, $\mc{N}=2$, $\mr{U}(N)$ gauge theory with $N_{\mr{F}}$ fundamental hypermultiplets; it becomes superconformal if $N_{\mr{F}}=2N$. }
  \label{fig:2}
\end{figure}

Let us now discuss the interesting situation depicted in fig.~\ref{fig:2}.
\footnotetext{All the multiplets here are $\mc{N}=(2,2)$ multiplets.}

\noindent In this case we have an extra holonomy coming from the left-moving $\mr{R}$-charge. Since the theory has not superpotential, we are free to chose the $\mr{R}$-charge of each multiplet: we will take all of them as zero. The holonomies are listed in tab.~\ref{tab:22}.

\begin{table}[H]
 \centering
 \begin{tabular}{lc}
  \toprule
  Group & Fugacity\\
  \midrule
  $\mr{U}(k)$ & $u_i$\\
  $\mr{U}(N)$ & $z_\alpha$\\
  $\mr{U}(N_\mr{F}-N)$ & $\mu_A$\\
  $\mr{U}(1)_\Phi$ & $-\epsilon$\\
  $\mr{U}(1)_R$ & $\ell$\\
  \bottomrule
 \end{tabular}
 \caption{Holonomies and Fugacities of the $\mc{N}=(2,2)$ case.}\label{tab:22}
\end{table}
\noindent As before, these fugacities are not independent:
\begin{equation}
 \sum_{\alpha=1}^Nz_\alpha + \sum_{A=1}^{N_{\mr{F}}-N}\mu_A = 0.
\end{equation}
Then the elliptic genus reads:
\begin{equation}
 \mc{Z}_k^{(2,2)}(\vec{z}, \vec{\mu}, \epsilon, \ell, \tau) =
 \int Z_{\rm{vec}}(\vec{u},\ell, \tau) Z_\Phi(\vec{u},\epsilon, \ell, \tau) Z_I(\vec{u}, \vec{z}, \ell, \tau) Z_J(\vec{u}, \vec{\mu} ,\ell, \tau),
\end{equation}
where the contributions of the multiplets are respectively:
\begin{align}
  Z_{\rm{vec}}(\vec{u},\ell, \tau) &= \f{1}{k!}\l(-\f{2\pi\eta^3(q)}{\theta_1(\tau | \ell)}\r)
 \prod_{\substack{r,s=1\\ r\neq s}}^k \f{\theta_1(\tau | u_r - u_s)}{\theta_1(\tau | u_r - u_s -\ell)}
 \prod_{r=1}^k \d u_r,\\
 Z_\Phi(\vec{u}, \epsilon, \ell, \tau) & = \label{eq:2B}
 \prod_{r,s=1}^k \f{\theta_1(\tau | u_r - u_s - \ell - \epsilon)}{\theta_1(\tau | u_r - u_s - \epsilon)},\\
 Z_I(\vec{u}, \vec{z}, \ell, \tau) &= \label{eq:2I}
 \prod_{\alpha=1}^N \prod_{r=1}^k \f{\theta_1(\tau | u_r - \ell - z_\alpha)}{\theta_1(\tau | u_r - z_\alpha)},\\
 Z_J(\vec{u}, \vec{\mu}, \ell, \tau) &= \prod_{A}^{N_{\mr{F}}-N}\prod_{r=1}^k \f{\theta_1(\tau | -u_r -\ell +\mu_A)}{\theta_1(-u_r + \mu_A)}.
\end{align}
Notice that a different $\mr{R}$-charge assignment would lead to the same partition function, provided we reabsorb various $\ell^{\f{R}{2}}$ factors in $\mr{U}(1)_\Phi$ and in a global $\mr{U}(1)$ subgroup of $\mr{U}(N)$ and $\mr{U}(N_{\mr{F}}-N)$.

\subsubsection{Polology}
The discussion about poles in the present case resembles that of paragraph \ref{sss:polology1}: here we will spot the differences with the previous case. The hyperplanes are:
\begin{align}
  H_{\mr{vec};r,s} &= \l\{ u_r - u_s -\ell =0 \r\},
  & H_{\Phi;r,s} &= \l\{ u_r - u_s - \epsilon =0 \r\},\\
  H_{I;r\alpha} &= \l\{ u_r = z_\alpha \r\},
  & H_{J;r,A} &= \l\{ u_r = \mu_A \r\}.
\end{align}
As before poles coming from $H_J$ are discarded by JK procedure. We will organize the contributions as before: fixing a color we have that the first pole occurs at $u_{\alpha,1}= z_\alpha$. To compute the second pole of that color, we can use either $H_\mr{vec}$ or $H_\Phi$; in the first case we have $u_{\alpha,2} = z_\alpha + \ell$. This value of $u$ indeed, set the numerator of eq. \eqref{eq:2I} to zero and so we drop this case; the remaining choice is to take $u_{\alpha,2}=z_\alpha + \epsilon$. Then we proceed to the third pole of the same color: as before, we can have either $u_{\alpha,3} = z_\alpha + \epsilon + \ell$ or $z_{\alpha,3} = z_\alpha + 2\epsilon$. Again, in the first case we encounter a zero in the numerator of eq. \eqref{eq:2B}. The iteration of this argument shows that the poles are of the form:
\begin{equation}\label{eq:2poles}
 u_{\alpha,r_\alpha} = z_\alpha + (r_\alpha -1)\epsilon,
\end{equation}
with $r_\alpha\in [1,k_\alpha]$ and $\sum_{\alpha} k_\alpha =k$. We conclude that the pole structure is the same of that of the $\mc{N}=(0,2)$ theory.

\subsubsection{Computation}
As in the previous case, we have just to plug the poles \eqref{eq:2poles} into the expression of one loop determinants of the multiplets and to use eq. \eqref{eq:theta_res}:
\begin{align}
 \mc{Z}_{\mr{vec}} &= 
  \l(\f{2\pi\eta^3(q)}{\theta_1(\tau | -\ell)} \r)^k
  \prod_{\substack{\alpha,\beta=1 \\ \alpha\neq\beta}}^N
  \prod_{r_\alpha=1}^{k_\alpha} \prod_{s_\beta=1}^{k_\beta}
  \f{\theta_1(\tau | z_{\alpha\beta} + (r_\alpha - s_\beta)\epsilon)}{\theta_1(\tau | z_{\alpha\beta}+(r_\alpha -s_\beta)\epsilon - \ell)}\times\nonumber\\
  &\hspace{2cm}\times
  \prod_{\alpha=1}^N\prod_{\substack{r_\alpha,s_\alpha=1\\ r_\alpha\neq s_\alpha}}^{k_\alpha}
  \f{\theta_1(\tau | (r_\alpha - s_\beta)\epsilon)}{\theta_1(\tau | (r_\alpha -s_\beta)\epsilon - \ell)},\\
 \mc{Z}_\Phi &=
 \prod_{\substack{\alpha,\beta=1 \\ \alpha\neq\beta}}^N
   \prod_{r_\alpha=1}^{k_\alpha} \prod_{s_\beta=1}^{k_\beta}
   \f{\theta_1(\tau |z_{\alpha\beta} + (r_\alpha - s_\beta-1)\epsilon-\ell )}
   {\theta_1(\tau |z_{\alpha\beta} + (r_\alpha - s_\beta-1)\epsilon )}\times\nonumber\\
   &\hspace{2cm}\times
   \prod_{\alpha=1}^N
   \l(\f{\theta_1(\tau|-\ell)}{2\pi\eta^3(q)} \r)^{k_\alpha-1}
   \prod_{\substack{r_\alpha,s_\alpha=1\\ r_\alpha\neq s_\alpha+1}}^{k_\alpha}
   \f{\theta_1(\tau |(r_\alpha - s_\beta-1)\epsilon-\ell )}
   {\theta_1(\tau |(r_\alpha - s_\beta-1)\epsilon )},\\
  \mc{Z}_I &=
  \prod_{\substack{\alpha,\beta=1 \\ \alpha\neq\beta}}^N
  \prod_{r_\alpha=1}^{k_\alpha}
  \f{\theta_1(\tau| z_{\alpha\beta} +(r_\alpha-1)\epsilon-\ell)}
  {\theta_1(\tau | z_{\alpha\beta} +(r_\alpha-1)\epsilon)}\times
  \prod_{\alpha=1}^N
  \l(\f{\theta_1(\tau | -\ell)}{2\pi\eta^3(q)}\r)  
  \prod_{r_\alpha=2}^{k_\alpha}
  \f{\theta_1(\tau|(r_\alpha-1)\epsilon-\ell)}
  { \theta_1(\tau |(r_\alpha-1)\epsilon)},\\
  \mc{Z}_J &=
  \prod_{\alpha,\beta=1}^N
  \prod_{r_\alpha=1}^{k_\alpha}
  \f{\theta_1(\tau | \mu_\alpha - z_\beta -(r_\alpha-1)\epsilon-\ell)}
  {\theta_1(\tau | \mu_\alpha - z_\beta -(r_\alpha-1)\epsilon)}.
\end{align}
Using again eqs. \eqref{eq:simp1} and \eqref{eq:simp2} both for numerators and denominators, and collecting all the factors, it is possible to write the result as:
\begin{equation}\label{eq:fin22}
\mc{Z}_{\vec{k}}^{(2,2)}(\vec{z}, \vec{\mu},\epsilon,\ell, \tau) = 
\prod_{\alpha,\beta=1}^N
\f{\Theta_1(\tau,\epsilon | z_{\alpha\beta}-k_\beta\epsilon-\ell)_{k_\alpha}}
{\Theta_1(\tau,\epsilon | z_{\alpha\beta}-k_\beta\epsilon)_{k_\alpha}}
\f{\Theta_1(\tau,\epsilon |z_\beta-\mu_\alpha+\ell)_{k_\alpha}}
{\Theta_1(\tau,\epsilon |z_\beta-\mu_\alpha)_{k_\alpha}},
\end{equation}
which, unfortunately, has not a simple resummation in terms of elliptic hypergeometric equation, due to the presence of the shift $-\ell$ in the numerator of the first factor. This result reproduces holomorphic 4d blocks and free-field correlators of elliptic Virasoro algebrae \cite{nieri}. Eq.~\eqref{eq:fin22} can be used to generalize the result for the topological vertex presented in \cite{vortex} to its elliptic version \cite{hiv}, in which rotational modes are resummed. Following \cite{vortex} we define the ``elliptic-lift'' of the amplitude of the open topological string as a double copy of our original system with opposite $\ell$:
\begin{equation}\label{eq:amplit}
\mc{A}^{\mr{ell.}}_N(\vec{z}, \vec{\mu},\epsilon,\ell, \tau)=\sum_{k_1=1}^\infty\dots\sum_{k_N=1}^\infty \mc{Z}_{\vec{k}}^{(2,2)}(\vec{z}, \vec{\mu},\epsilon,\ell, \tau)
\mc{Z}_{\vec{k}}^{(2,2)}(\vec{z}, \vec{\mu},\epsilon,-\ell, \tau)
z^{|\vec{k}|}.
\end{equation}
Now we can apply, as before, eq. \eqref{eq:trick} in the denominator as well as in the numerator of \eqref{eq:amplit}. Using our bag of tricks as before the result reads:
\begin{equation}
\mc{A}^{\mr{ell.}}_N(\vec{z}, \vec{\mu},\epsilon,\ell, \tau)=
  \mc{D}^{(2,2)}_N
  \prod_{\alpha=1}^N\rEs{4N}{4N-1}{\vec{A}_\alpha^+, \vec{A}_\alpha^-, \vec{B}_\alpha^+, \vec{B}_\alpha^-}{\epsilon, \vec{C}_\alpha , \vec{C}_\alpha,\vec{D}_\alpha,\vec{D}_\alpha}{\tau,\epsilon}{z_\alpha},
\end{equation}
where we set:
\begin{align}
 \vec{A}_\alpha^\pm &= \l(z_\alpha - z_1 \pm\ell + \epsilon,\dts,z_\alpha - z_N \pm\ell + \epsilon  \r);\\
 \vec{B}_\alpha^\pm &= \l(z_1 - \mu_\alpha \pm \ell,\dts, z_N - \mu_\alpha \pm \ell \r);\\
 \vec{C}_\alpha &= \l(z_\alpha - z_1,\dts, \widehat{z_\alpha - z_\alpha},\dts, z_\alpha - z_N \r);\\
 \vec{D}_\alpha &= \l( z_1 - \mu_\alpha,\dts,z_N - \mu_\alpha \r);\\
 \mc{D}^{(2,2)}_N &= 
 \prod_{1\leq\alpha < \beta\leq N}e^{i\pi \epsilon(z_\alpha\partial_{z_\alpha}-z_\beta \partial_{z_\beta})}
 \f{%
 \theta(\tau | z_{\alpha\beta}-\ell)\theta(\tau | z_{\alpha\beta} +\ell)
 }{%
 \theta^2(\tau | z_{\alpha\beta})
 }\times\nonumber\\
 &\hspace{2cm}
 \times
 \f{%
 \theta^2(\tau | z_{\alpha\beta}+\epsilon(z_\alpha\partial_{z_\alpha} - z_\beta\partial_{z_\beta}))
 }{%
 \theta(\tau |z_{\alpha\beta}+\ell+\epsilon(z_\alpha\partial_{z_\alpha} - z_\beta\partial_{z_\beta}) )
  \theta(\tau |z_{\alpha\beta}-\ell+\epsilon(z_\alpha\partial_{z_\alpha} - z_\beta\partial_{z_\beta}) )
 }.
\end{align}
This result is the elliptic analogue of the $\mc{N}=2$ vortex partition presented in \cite{vortex}.

\section{Conclusion and outlook}

In the present work we presented a detailed derivation of elliptic vortex partition function of respectively $\mc{N}=1$ and $\mc{N}=2$ gauge theories, using the machinery of elliptic genus, developed in \cite{bh}, and resumming over all vortices configurations. In the former case we made an elliptic generalization of the vortex partition function for fixed $k$ contained in \cite{bc,gd}; furthermore we were able to resum all the contribution against a vorticity parameter, giving a generalization to $\mr{U}(N)$ vortices of the result presented in \cite{hf4}. In the latter we extended the result found in \cite{vortex} to the elliptic case.

A couple of remarks are in order about possible extensions of the present analysis.
 The set up of this paper can be used to analyze the elliptic
 genus of other target spaces, providing information on their elliptic quantum cohomology \cite{ao}.
 Moreover, one could extend the present framework to other geometries such as product spaces 
$\mb{C}\times\Sigma$ and total spaces of line bundles over a Riemann surface, such as $T^*\Sigma$, hopefully helping to clarify the algebraic
structures of BPS vacua of gauge theories on these manifolds.

\acknowledgments We would like to thank F. Benini, G. Bonelli and A. Tanzini for suggesting me this topic and for precious discussions. This research is partially supported by the INFN i.s. ST\&{}FI.

\appendix

\section{Special Functions}

All through the paper several special function are used. Here we list them with some useful properties.

\subsection{$q$-Pochhammer, $\eta$ and $\theta$}

First of all we define the modular parameter to be $q=e^{2\pi i \tau}$, with $\Im(\tau)>0$. The $q$-Pochhammer is defined:
\begin{equation}
 (y,q)_\infty := \prod_{k=0}^\infty(1-yq^k).
\end{equation}
In term of this function is possible to define the Dedekind eta:
\begin{equation}
 \eta(q):= q^{\f{1}{24}}(q;q)_\infty.
\end{equation}
and the ``core'' of Jacobi theta functions:
\begin{equation}
 \theta(\tau | z):= (y;q)_\infty (qy^{-1};q)_\infty,
\end{equation}
where we set for convenience $y=e^{2\pi i z}$. The most ubiquitous function in this paper is the Jacobi theta of first kind:
\begin{equation}
 \begin{split}
  \theta_1(\tau | z) &:= i q^{\f{1}{8}}y^{-\f{1}{2}}(q;q)_\infty \theta(\tau | z)\\
  & = -i q^{\f{1}{8}} y^{\f{1}{2}}(q;q)_\infty \theta(\tau | -z).
 \end{split}
\end{equation}
From its definition is possible to see that Jacobi function is odd:
\begin{equation}
 \theta_1(\tau | -z) = -\theta_1(\tau | z).
\end{equation}
Moreover, under shift of the argument $z\mapsto z+a+b\tau$ ($a,b\in\mb{Z}$), the function transforms as:
\begin{equation}\label{eq:thetatrasf}
 \theta_1(\tau | z + a + b\tau) = (-1)^{a+b}e^{-2\pi i b z}e^{-i\pi b^2\tau}\theta_1(\tau | z).
\end{equation}
So we see that it is $1$-periodic and $\tau$-quasiperiodic. Then we can reduce to study its behaviour in a ``fundamental domain'' of the lattice $\mb{Z}+\tau\mb{Z}$. The function $\theta_1(\tau | z)$ has no poles and simple zeroes occur at $z=\mb{Z}+\tau\mb{Z}$. The residues of its inverse are
\begin{equation}\label{eq:theta_res}
 \f{1}{2\pi i}\oint_{z=a+b\tau}\f{\d z}{\theta_1(\tau | z)} = \f{(-1)^{a+b}e^{i\pi b^2 \tau}}{2\pi \eta^3(q)};
\end{equation}
and for small values of $q$ and $z$ we have:
\begin{equation}\label{eq:theta_small}
 \theta_1(\tau | z) \xrightarrow{q\rightarrow 0} 2q^{\f{1}{8}}\sin(\pi z) \xrightarrow{z\rightarrow 0} 2\pi q^{\f{1}{8}}z.
\end{equation}
\subsection{$\Theta$ and Elliptic Hypergeometric Functions}\label{app:Theta}
We give the following definitions which is suitable for our results:
\begin{equation}
 \Theta_\bullet(\tau, \sigma | a)_n := 
 \begin{cases}
 \displaystyle\prod_{k=0}^{n-1}\theta_\bullet(\tau | a + k\sigma) & n\in \mb{Z}_\geq,\\
 \l[\displaystyle\prod_{k=0}^{|n|-1}\theta_\bullet(\tau | a - (k+1)\sigma)\r]^{-1} & n\in\mb{Z}_<;
 \end{cases}
\end{equation}
where $\theta_\bullet$ can be just $\theta$ or $\theta_1$, and the same for $\Theta_\bullet$.
This function enjoys the key property of Pochhammer symbol:
\begin{equation}
    \Theta_\bullet(\tau, \sigma| a)_m = \Theta_\bullet(\tau,\sigma |a)_n\Theta_\bullet(\tau,\sigma| a+n\sigma)_{m-n}.
\end{equation}
Using eq. \eqref{eq:theta_small}, indeed, it is possible to show that for small values of its arguments:
\begin{equation}
 \Theta(\tau,\sigma | a) \xrightarrow[a+n\sigma \rightarrow 0]{q\rightarrow 0} [-2\pi i\sigma]^n\left(\f{a}{\sigma} \right)_n;
 \qquad\qquad\quad
 \Theta_1(\tau,\sigma | a) \xrightarrow[a+n\sigma \rightarrow 0]{q\rightarrow 0} \l[2\pi q^{\f{1}{8}}\sigma\r]^n\left(\f{a}{\sigma} \right)_n.
\end{equation}
Another identity which will be useful is the following which holds\footnote{This is actually a property that $\theta_1$ and its products share with all odd functions.} just for $\Theta_1$:
\begin{equation}\label{eq:trick}
 \Theta_1(\tau,\sigma | a - l\sigma) \Theta_1(\tau,\sigma | -a -m\sigma) = \f{\theta_1(\tau | a)}{\theta_1(\tau | a + (m-l)\sigma)} \Theta_1(\tau,\sigma | a+ \sigma)_m \Theta_1(\tau,\sigma |-a+\sigma)_l.
\end{equation}
We shall use $\Theta$ to write products of $\Theta_1$. In the present cases we have the following situation:
\begin{equation}\label{eq:baltheta}
 \f{
 \prod_{i=1}^M \Theta_1(\tau,\epsilon | z_i)_n
 }{
 \prod_{i=1}^M \Theta_1(\tau,\epsilon | a_i )_n
 }
 =
 e^{i\pi n\l(\sum_{i=1}^M a_i - \sum_{i}^M z_i \r)}
 \f{
 \prod_{i=1}^M \Theta(\tau, \sigma | z_i )_n
 }{
 \prod_{i=1}^M \Theta(\tau, \sigma | a_i )_n
 }.
\end{equation}
Elliptic hypergeometric series \cite{spiridonov} are defined as follows:
\begin{equation}
 \rEs{r}{s}{t_0,\dts,t_{r-1}}{w_1,\dts,w_s}{\tau,\sigma}{z}:=
 \sum_{k=0}^\infty \f{%
 \prod_{i=0}^{r-1}\Theta(\tau,\sigma| t_i)_k
 }{%
 \Theta(\tau,\sigma | \sigma)_k\prod_{j=1}^{s}\Theta(\tau,\sigma| w_j)_k
 }
 z^k.
\end{equation}
A remarkable property is the following\footnote{Similar operators appear in \cite{frv}.}:
\begin{equation}\label{eq:derE}
 \theta(\tau | b + z\partial_z) \rEs{r}{s}{t_0,\dts,t_{r-1}}{w_1,\dts,w_s}{\tau,\sigma}{z}
 =
 \theta(\tau| b)\rEs{r+1}{s+1}{t_0,\dts,t_{r-1},b+\sigma}{w_1,\dts,w_s,b}{\tau,\sigma}{z}.
\end{equation}
In the case of $b=0$, eq. \eqref{eq:derE} reduces to:
\begin{equation}
 \theta(\tau |z\partial_z) \rEs{r}{s}{t_0,\dts,t_{r-1}}{w_1,\dts,w_s}{\tau,\sigma}{z}
 =
 \f{%
 \prod_{i=0}^{r-1}\theta(\tau | t_i)%
 }{%
 \prod_{j=1}^{s}\theta(\sigma | w_i)%
 }
 \rEs{r}{s}{t_0+\sigma,\dts,t_{r-1}+\sigma}{w_1+\sigma,\dts,w_s+\sigma}{\tau,\sigma}{z};
\end{equation}
which resembles the usual relation between $\tensor[_r]{F}{_s}$ and its derivatives.

\bibliography{bibliography}
 
\end{document}